\documentclass[%
 reprint,
 amsmath,amssymb,
 aps,
prl,
floatfix,
]{revtex4-2}
 
\usepackage{graphicx}
\usepackage{dcolumn}
\usepackage{bm}
\usepackage{hyperref}

\begin{document}
 
 
\title{\textbf{Relaxation-Time Boltzmann Solutions Without Truncations} 
}%

\author{Tiago Nunes da Silva}
\email{Contact author: t.j.nunes@ufsc.br}
\author{Jadna L. Barauna}%
\affiliation{%
  Departamento de Física, Universidade Federal de Santa Catarina, Florianópolis, Brazil
}%

\author{Giorgio Torrieri}
\affiliation{%
 IFGW, Unicamp, Campinas, Brazil
}%

 
\date{\today}
 
\begin{abstract}
The relaxation-time Boltzmann equation governs relativistic matter far from equilibrium, but existing approaches rely on finite-order truncations that break down in this regime. We introduce ShARK (Stochastic Advection Relaxation Kinetics), the first fully 3D Monte Carlo kinetic framework to capture exact, far-from-equilibrium ultrarelativistic dynamics for arbitrary initial conditions while directly enforcing target transport coefficients, converging to the exact Anderson-Witting solution. Validated against the Bjorken and Gubser flows, the framework reproduces the exact kinetic result where truncated methods fail.
\end{abstract}
 
\maketitle
 
 
The relativistic Boltzmann equation in the relaxation-time approximation (RTA) governs the microscopic dynamics of matter far from local equilibrium, from the earliest, most strongly gradient-dominated stages of a heavy-ion collision to neutrino transport in the early Universe and the ringdown of compact merger remnants~\cite{Anderson:1974nyl, deGroot:1980dk, Hannestad:1995rs, Romatschke:2017ejr, Alford:2017rxf, Kurkela:2018wud, Foucart:2020qjb, Froustey:2020mcq}. Despite its central role, no existing numerical method solves the RTA Boltzmann equation exactly and generally in the ultrarelativistic, far-from-equilibrium regime where these applications live.
 
Three classes of methods are currently in use to approximate the collision term, each limited in a distinct way. Relativistic dissipative hydrodynamics (Israel-Stewart, DNMR) truncates the infinite moment hierarchy of the Boltzmann equation at second order in a gradient expansion around local equilibrium~\cite{Israel:1979wp, Denicol:2012cn}; by construction, this truncation captures only the leading deviations from equilibrium and is not guaranteed to converge as the expansion parameter grows large, precisely the regime that governs early-time heavy-ion dynamics. Relativistic lattice Boltzmann (LB) solvers instead construct the distribution function from a finite-order expansion on orthogonal polynomials, discretized via Gauss quadrature~\cite{Mendoza:2010as, Romatschke:2011qp, Romatschke:2011hm, BAZZANINI2021101320, Gabbana:2019ydb}; this construction is likewise formulated in the limit of small Knudsen number and becomes increasingly inaccurate as the system departs from local equilibrium~\cite{BAZZANINI2021101320}, and extending LB toward the free-streaming limit remains an active area of development. Parton cascade models~\cite{Zhang:1997ej, Zhang:1999bd,Xu:2004mz}, in contrast, resolve the full nonlinear collision term through explicit microscopic scattering, but their transport coefficients emerge indirectly from the underlying cross sections, making direct, controlled comparison to a target $(\eta/s)(T)$ difficult in practice.
 
\begin{figure}[t]
    \centering
    \includegraphics[width=\linewidth]{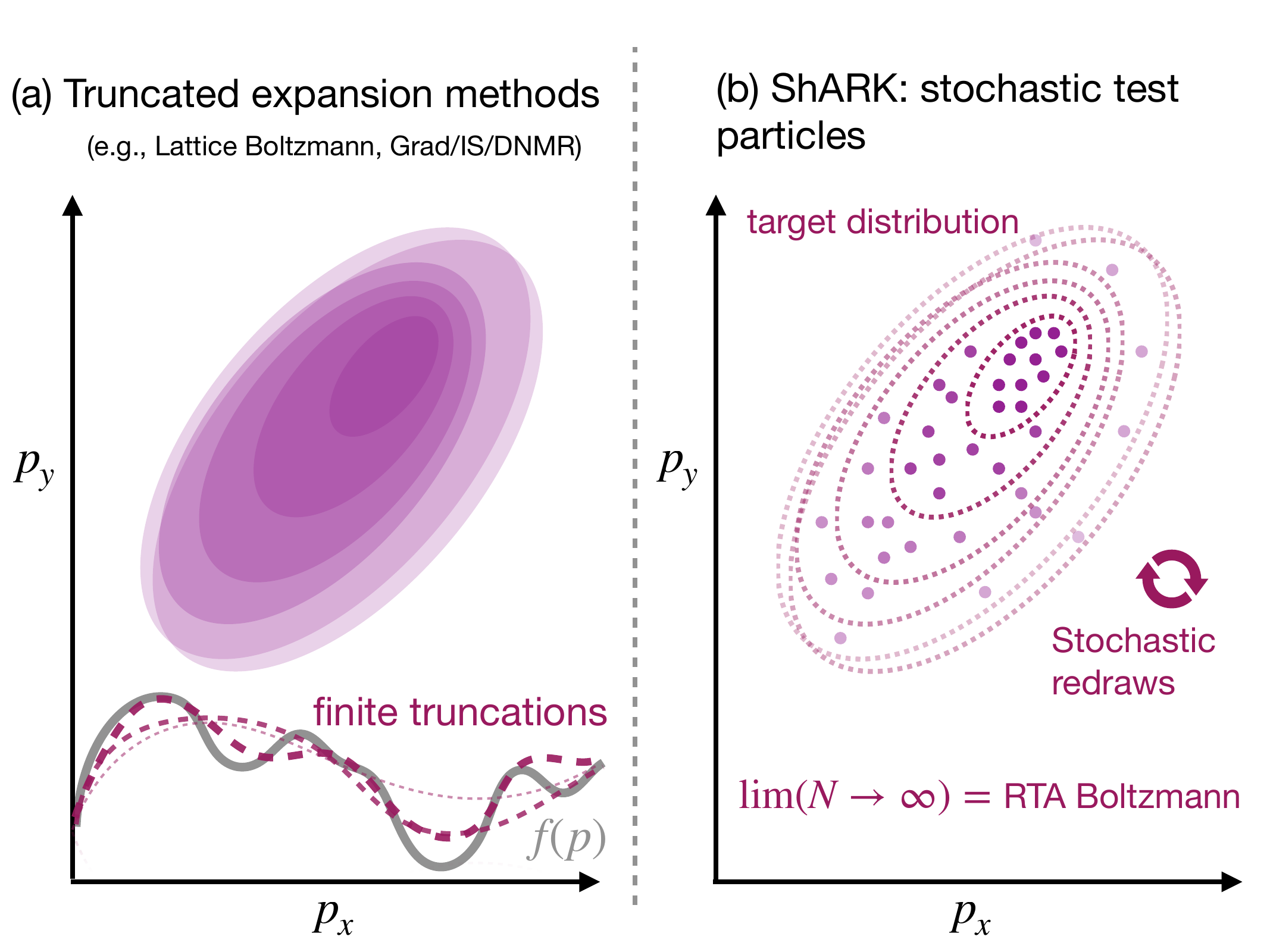}
    \caption{Schematic contrast between truncation-based and ShARK representations of the one-particle distribution function. \textbf{(a)} Truncation-based methods (lattice Boltzmann, hydrodynamic moment expansions) represent $f(x,p)$ in a finite basis of orthogonal polynomials or moments; the mode retained at a given truncation order (solid) captures the leading behavior near local equilibrium, while a higher-order mode (faded, dashed) that encodes genuine far-from-equilibrium structure is discarded. \textbf{(b)} ShARK represents $f(x,p)$ directly as an ensemble of $N$ test particles (purple) undergoing exact microcanonical relaxation events that conserve the cell's total invariant mass $M$ and particle number $N$ at every collision, with no truncation order and no discarded modes.}
    \label{fig:schematic}
\end{figure}
 
Here we introduce ShARK (Stochastic Advection Relaxations Kinetics), a $3+1$D stochastic framework to solve the Anderson-Witting RTA equation that avoids these limitations (Fig.~\ref{fig:schematic}). Rather than truncating the distribution function in momentum space, ShARK represents it directly via an ensemble of test particles undergoing exact, microcanonical relaxation events. The evolution follows an advection-relaxation splitting: particles free-stream between time steps, and each fluid cell's momenta are then stochastically relaxed. To match a target shear viscosity $\eta/s$, the local relaxation time is dynamically determined via the Chapman-Enskog relation, $\tau_R = 5(\eta/s)/T$~\cite{ANDERSON1974489, Florkowski:2013lya}. The relaxation step is triggered according to the Poisson probability $P_{\rm trig} = 1 - \exp(-\Delta t / \gamma_{\rm cell} \tau_R)$, natively accounting for relativistic time dilation via the cell's Lorentz factor $\gamma_{\rm cell}$. This construction is conceptually akin to lattice Boltzmann, but carried out in continuous momentum space rather than on a fixed, discretized momentum set. We show (see End Matter and the companion paper~\cite{ShARK_Long}) that this construction converges, at large particle multiplicity $N$, to the exact solution of the Anderson-Witting equation, with no truncation order to break down. We validate this claim against the two symmetry-reduced systems for which exact kinetic-theory solutions exist -- the Bjorken~\cite{Bjorken:1982qr} and Gubser~\cite{Gubser:2010ze} flows -- and demonstrate that ShARK tracks the exact solution deep into regimes where second-order hydrodynamic truncations (Israel-Stewart, DNMR) visibly break down, the same failure mode that limits truncation-based kinetic approaches more broadly.
 
The microscopic mechanism underlying this convergence is the collision kernel itself. Within each fluid cell, ShARK relaxes the local distribution via the RAMBO algorithm~\cite{Kleiss:1985gy}, originally developed to sample final-state phase space in high-energy scattering: given the cell's total four-momentum, RAMBO redraws a new set of $N$ on-shell test-particle momenta uniformly from the Lorentz-invariant phase space, exactly conserving energy and momentum at every event. Treating repeated redraws as a Markov process on the $N$-particle phase space, a molecular-chaos closure reduces this exactly to a one-body master equation whose full-redraw transition rate factorizes, in the continuum limit, into the Anderson-Witting collision operator. This constitutes a relativistic physical adaptation of the rigorous stochastic-to-continuum limit recently established for the non-relativistic BGK equation by Buttà and Pulvirenti~\cite{butta2023particlesystemsbgkequation}. Thus, the construction not only recovers the correct equilibrium statistics but also reproduces the exact non-equilibrium relaxation dynamics at rate $1/\tau_R$. As a check on the equilibrium limit of this process: for any finite $N$, the single-particle energy spectrum generated by a single redraw is known in closed form as a finite-size microcanonical distribution, and we show that it converges pointwise to the canonical J\"uttner-Boltzmann distribution as $N\to\infty$. Local equilibrium is thus never imposed; it emerges as the large-$N$ limit of a genuinely microcanonical process. We give the full derivation of this master-equation reduction, together with its numerical verification, in the End Matter.

Because the local test-particle occupancy scales with the expanding fluid's energy density, this finite-$N$ phase space provides a natural, parameter-free interpolation between dynamical regimes. By defining an oversampling parameter $\alpha$ that sets the number of test particles per unit entropy, the system transitions continuously from a dense hydrodynamic core ($N \gg 1$) to a stochastic kinetic regime. Crucially, as peripheral or late-time cells expand and their occupancy drops below $N = 2$, the microcanonical redraw can no longer simultaneously conserve energy and momentum. Collisions natively cease, and the local dynamics seamlessly freeze out into exact ballistic free-streaming. This local, geometry-sensitive decoupling removes the need for a global switching hypersurface between hydrodynamic and kinetic-transport descriptions, as used in hybrid models~\cite{Petersen:2008dd}. Converting the oversampled partonic test-particle ensemble into physical, decaying hadrons at freeze-out remains a distinct physical process deferred to future work.

To validate the collision kernel and isolate it from spatial Cartesian-grid artifacts, we map both expanding geometries into static computational boxes, following the approach used in parton cascade codes such as BAMPS~\cite{Xu:2004mz}. In this homogeneous limit, the computational grid acts strictly as an ensemble of independent statistical realizations rather than a physical volume. The macroscopic volume expansion is implemented microscopically via an exact continuous redshift of particle momenta during the advection phase, allowing spatial profiles to be evaluated at arbitrary Minkowski radii via direct coordinate transformations of the temporal state and circumventing finite-size boundary effects.

As a first test of the collision kernel, we consider the boost-invariant, transversely homogeneous Bjorken expansion~\cite{Bjorken:1982qr}. ShARK reproduces the exact ideal-fluid cooling law $T\propto\tau^{-1/3}$, and the expected viscous-heating correction at finite $\eta/s$, in the companion paper~\cite{ShARK_Long}. Because Bjorken flow constrains only the time dependence of a spatially uniform system, it exercises the relaxation kernel without engaging the advection step. This provides a clean, isolated test of whether the kernel alone reproduces the collision integral's known non-equilibrium attractor, which we present in the End Matter (Fig.~\ref{fig:bjorken_attractor}).

\begin{figure*}[t]
    \centering
    \includegraphics[width=\textwidth]{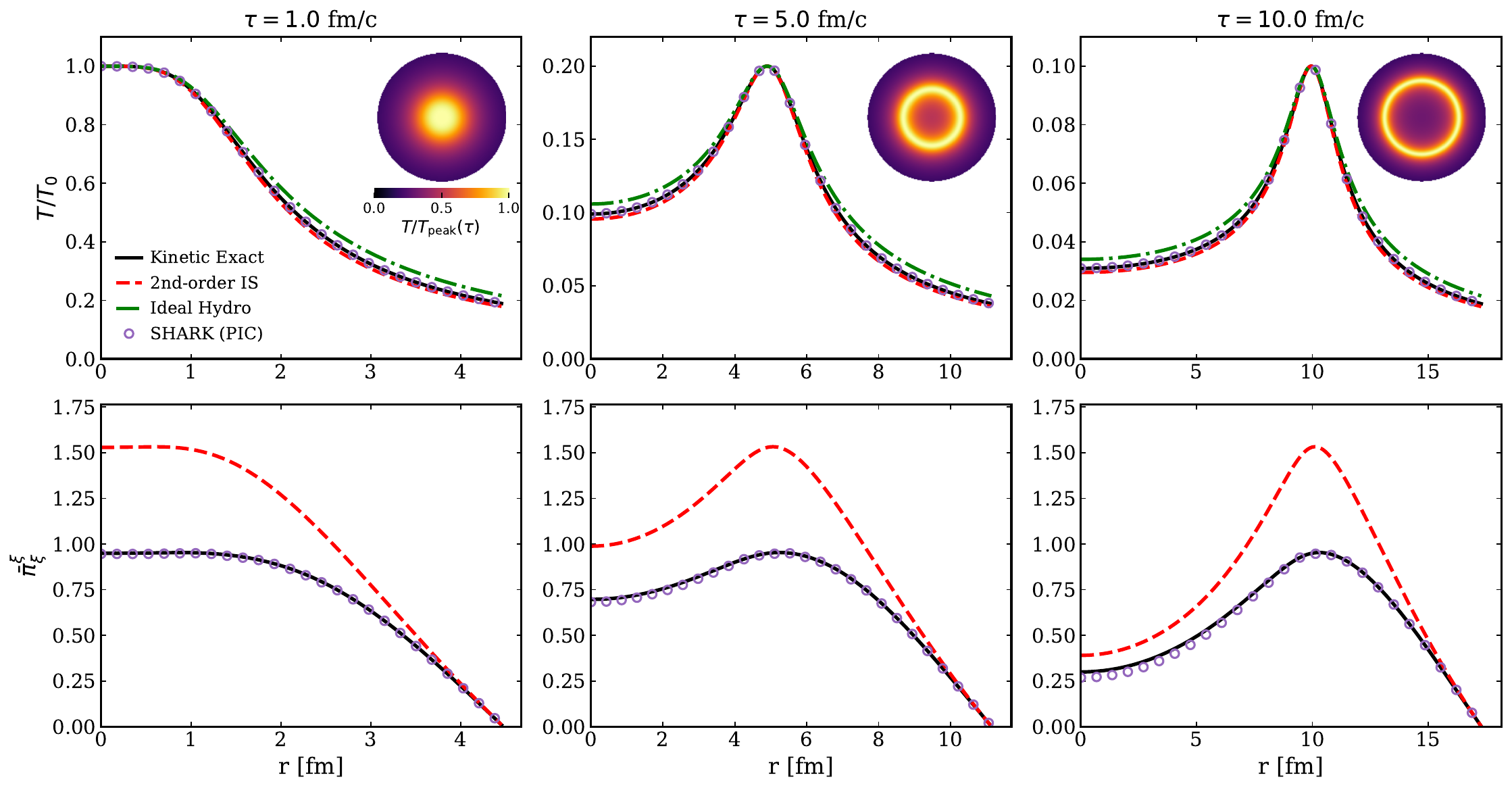}
    \caption{Minkowski radial profiles of the normalized temperature $T/T_0$ (top) and the shear-stress magnitude $\bar\pi^\xi_\xi$ (bottom) for the Gubser expansion at $4\pi\eta/s = 100$, shown at proper times $\tau = 1, 5$, and $10~\mathrm{fm}/c$. Insets (top) display 2D heatmaps of the transverse temperature profile, with the shared color scale normalized to the local peak temperature $T/T_{\mathrm{peak}}(\tau)$. The numerical ShARK solution (purple circles) is compared to ideal hydrodynamics (dash-dotted green), second-order Israel-Stewart theory (dashed red), and exact kinetic (solid black) solutions~\cite{Denicol:2014tha, Denicol:2014xca}}.
    \label{fig:gubser-profile}
\end{figure*}

A sharper test of exactness is the Gubser flow~\cite{Gubser:2010ze}, which improves on Bjorken flow by including genuine transverse expansion: its transverse and longitudinal expansion rates compete and invert in sign as the system evolves, driving large, spatially varying pressure anisotropies even at fixed shear viscosity. This makes Gubser flow the most demanding exact benchmark available for the collision kernel's accuracy across a continuously varying range of local gradient strength, independent of any spatial advection scheme. Its tractability follows from a hidden conformal symmetry, $SO(3)_q \otimes SO(1,1) \otimes \mathbb{Z}_2$, which reduces the Boltzmann equation to a macroscopic 0D temporal evolution in a rescaled de~Sitter time $\rho$, related to the physical coordinates via $\sinh\rho = -\left(1 - q^2\tau^2 + q^2r^2\right)/(2q\tau)$, with $q^{-1}$ setting the transverse size of the system. This single variable fully encodes the radial and temporal dependence of the flow shown in Fig.~\ref{fig:gubser-profile}.

We simulate the Gubser flow in a strongly dissipative regime characterized by a large shear viscosity, $4\pi\eta/s = 100$, and project the exact kinetic solution, the second-order Israel-Stewart (IS) approximation~\cite{Denicol:2014tha, Denicol:2014xca}, and the ShARK simulation back onto physical Minkowski coordinates $(\tau, r)$. Figure~\ref{fig:gubser-profile} shows the resulting radial profiles of the normalized temperature $T/T_0$ (top row) and the shear-stress magnitude $\bar\pi^\xi_\xi$ (bottom row) at three proper times spanning an order of magnitude in $\tau$. ShARK overlays the exact kinetic solution at every radius and time shown, correctly locating both the outward-propagating temperature peak and the node where the shear stress vanishes and the pressure anisotropy inverts sign. The IS approximation, in contrast, visibly departs from the exact solution: while it successfully tracks the macroscopic temperature profile, it substantially overestimates the shear-stress magnitude throughout the evolution. This inability to keep pace with the true dissipative gradients highlights the limits of truncation-based methods in far-from-equilibrium regimes.

\begin{figure}[t]
    \centering
    \includegraphics[width=\linewidth]{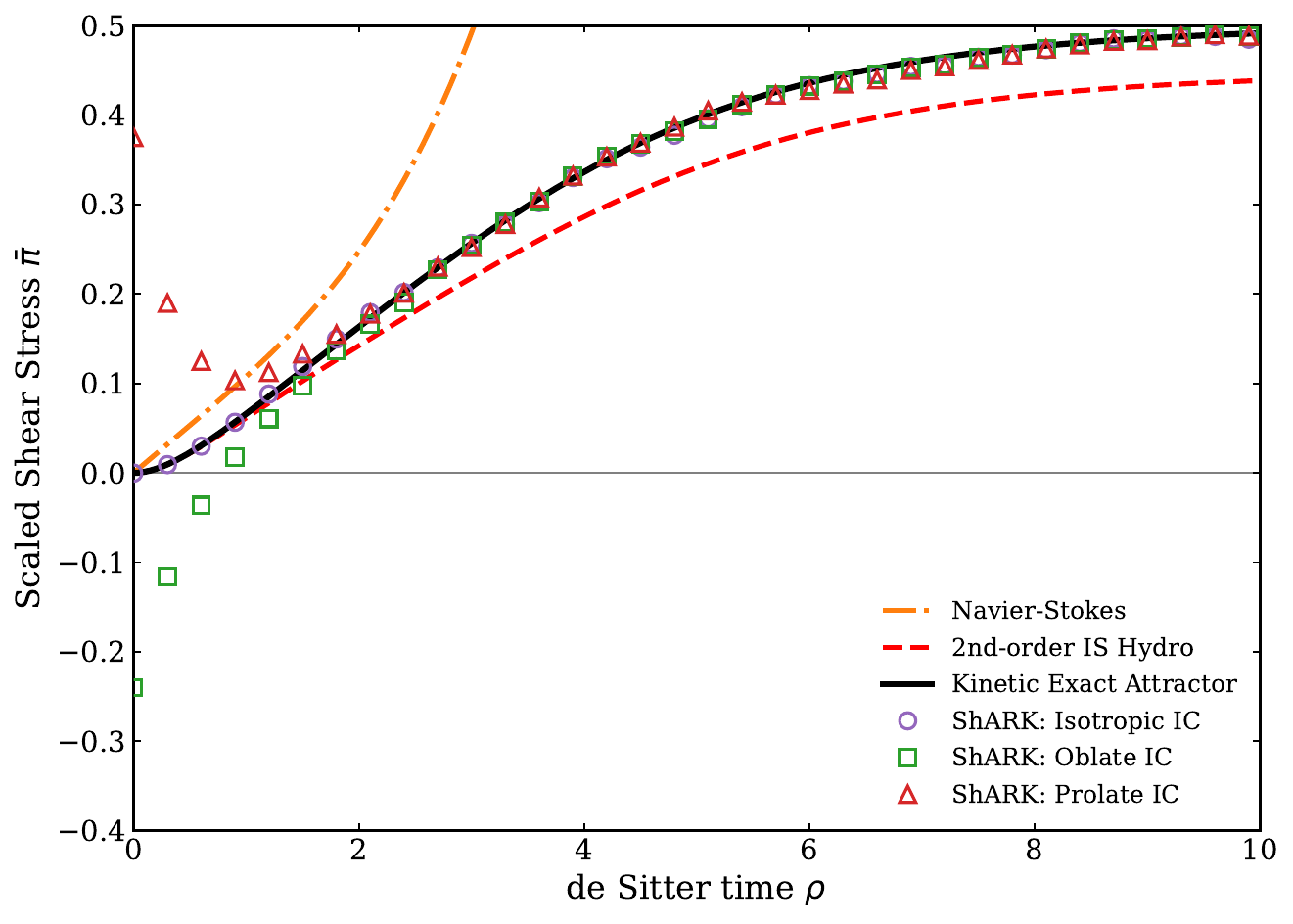}
    \caption{Non-equilibrium attractor of the scaled shear stress $\bar\pi(\rho)$ for the Gubser expansion at $4\pi\eta/s = 1$. ShARK simulations initialized in local equilibrium (purple circles), an extreme oblate state (green squares), and an extreme prolate state (red triangles) all collapse onto the exact kinetic attractor (solid black) within $\Delta\rho \approx 2$, while first-order Navier-Stokes (dash-dotted orange) and second-order Israel-Stewart theory (dashed red) deviate from it at large $|\rho|$~\cite{Denicol:2014tha, Behtash:2017wqg, Chattopadhyay:2018apf}.}
    \label{fig:gubser-attractor}
\end{figure}

A complementary, and in some sense more stringent, test is universality. Independent of its initial microscopic configuration, the true solution of the Boltzmann equation loses memory of its initial anisotropy and collapses onto a single non-equilibrium attractor curve $\bar\pi(\rho)$~\cite{Heller:2015dha, Behtash:2017wqg, Denicol:2018pak}. This is a genuine all-order statement about the infinite hierarchy of moments of the distribution function, and no finite truncation of that hierarchy guarantees it. Figure~\ref{fig:gubser-attractor} shows three independent ShARK simulations at $4\pi\eta/s = 1$, initialized respectively in local equilibrium ($\bar\pi_0 = 0$), in an extreme oblate configuration, and in an extreme prolate configuration. Despite these disparate initial conditions, all three trajectories converge onto the exact kinetic attractor obtained from the exact analytical solution of the RTA Boltzmann equation~\cite{Romatschke:2003ms, Denicol:2014tha, Behtash:2017wqg, Chattopadhyay:2018apf} within $\Delta\rho \approx 2$. Meanwhile, the first-order Navier-Stokes and second-order IS approximations shown for reference systematically miss the attractor at large $|\rho|$.
 
Together, these results show that ShARK does not merely improve truncation-based methods in accuracy; it removes truncation itself and reproduces the full, non-linear, all-order content of the Anderson-Witting equation using only local microcanonical relaxation, with no expansion order to break down as the system is driven arbitrarily far from equilibrium.
 
The construction underlying ShARK is not specific to heavy-ion collisions: it is a general algorithm for the ultrarelativistic Anderson-Witting equation, applicable wherever relaxation-time kinetics governs a system's approach to, or departure from, local equilibrium. Two examples illustrate this reach. In the remnant of a binary neutron star merger, neutrino transport crosses continuously within a single object from the optically thick, near-equilibrium core to the free-streaming exterior; because the local optical depth is direction- and time-dependent, the neutrinosphere is never a genuinely sharp surface, and state-of-the-art simulations instead rely on moment-closure (M1-type) schemes~\cite{Foucart:2020qjb} that introduce a fixed truncation order into the transport hierarchy. Cosmological neutrino decoupling presents an analogous problem in a homogeneous setting: as the weak interaction rate $\Gamma_\nu \sim G_F^2 T^5$ falls below the Hubble rate $H \sim T^2/M_{\rm Pl}$ near $T\sim 1\,{\rm MeV}$, the neutrino sector transitions continuously from tight coupling to collisionless free-streaming, a transition that standard Boltzmann solvers track by truncating the neutrino phase-space hierarchy at a finite multipole order~\cite{Froustey:2020mcq}. 

Like the spacetime switching surfaces used in heavy-ion models to patch together pre-equilibrium, fluid, and dilute hadronic stages, these momentum-space truncations share a fundamental conceptual limitation: they are ad-hoc interventions required because the macroscopic effective theory cannot natively span disparate kinetic regimes. However, the physical system does not select a sharp boundary between collisional and collisionless dynamics; it crosses continuously between regimes, locally and in time. Because ShARK solves the collision integral exactly rather than expanding around a fixed Knudsen number or multipole order, it degrades gracefully into an exact free-streaming solver in the dilute limit with no change to the underlying algorithm, extending its reach from the strongly coupled quark-gluon plasma to the weakly interacting regimes relevant to these astrophysical and cosmological settings. 

Across all these applications, the obstacle is the same one identified at the outset: existing approaches fix the complexity of the collision term in advance, whether by moment truncation, matched asymptotics, or an artificial switching surface (Fig.~\ref{fig:schematic}). ShARK removes this obstacle at its source, providing a common, truncation-free framework for far-from-equilibrium relativistic transport wherever the relaxation-time approximation applies.

\begin{acknowledgments}
The authors thank the members of the ExTrEMe Collaboration and R. Krupczak for fruitful discussions. The authors acknowledge the use of the Gemini AI model in code development. The authors are supported by CNPq through the INCT-FNA grant 408419/2024-5. T.N.dS. was supported by the Universal Grant 409029/2021-1. J.L.B. acknowledges a PhD fellowship from Fundacao de Amparo a Pesquisa e Inovacao do Estado de Santa Catarina (FAPESC) under grant 1947/2025. G.T. thanks Bolsa de produtividade CNPQ 305731/2023-8 and FAPESP temático 2023/13749--1 for support.
\end{acknowledgments}

\appendix

\section*{End Matter}
\label{sec:endmatter}

\subsection{Derivation of the Continuum Anderson-Witting Limit}

The RAMBO algorithm~\cite{Kleiss:1985gy} performs exact microcanonical sampling of the Lorentz-invariant phase space (LIPS). Using the Byckling-Kajantie recursive phase-space integration technique~\cite{BycklingKajantie}, the exact marginal single-particle energy spectrum for a finite number of massless particles $N$ in a cell with invariant mass $\sqrt{s}$ is given by~\cite{ShARK_Long}:
\begin{equation}
    f_{\rm micro}(E; N,s) \propto E\left(1 - \frac{2E}{\sqrt{s}}\right)^{N-3}
\end{equation}

For fixed $E$ and constant effective temperature $T_{\rm micro} = \sqrt{s}/(2N)$, taking the limit $N \to \infty$ exactly yields the canonical Jüttner-Boltzmann equilibrium distribution, $f_{\rm eq}(E) \propto E \exp(-E/T_{\rm micro})$.  

Because a full phase-space redraw replaces all momenta simultaneously with a configuration that retains no memory of the pre-collision state, the evolution of the joint $N$-particle distribution $F_N(\{\vec{p}\}, t)$ constitutes a Markov process governed by the $N$-body Master Equation:
  \begin{multline}
  \frac{\partial F_N(\{\vec{p}\},t)}{\partial t} = 
  \int d^{3N}p'\, \bigl[ \mathcal{W}_N(\{\vec{p}'\} \to \{\vec{p}\}) F_N(\{\vec{p}'\},t) \\
  {}- \mathcal{W}_N(\{\vec{p}\} \to \{\vec{p}'\}) F_N(\{\vec{p}\},t) \bigr].
\end{multline}
  
The uniform sampling of the microcanonical ensemble dictates that the transition rate is independent of the initial state, yielding $\mathcal{W}_N = (1/\tau_R) [dR_N(s) / \int dR_N(s)]$.  To close this hierarchy at the one-particle level, we invoke the molecular chaos hypothesis. Assuming that inter-particle correlations generated during a redraw are dissolved by spatial advection before the next collision step, the joint distribution factorizes prior to collision as $F_N \simeq \prod_{i=1}^N f_1(\vec{p}_i, t)$.  

Under molecular chaos, the surrounding $N-1$ particles act as an uncorrelated thermal bath, reducing the $N$-body transition rate to an effective one-body rate $W(\vec{p} \to \vec{p}')$. Because RAMBO performs a full redraw, the outgoing momentum is drawn independently of the initial state, so the transition rate factorizes as $W(\vec{p} \to \vec{p}') = \Gamma(\vec{p}')$.  Marginalizing the Master Equation over the unobserved $N-1$ momenta and substituting this factorized rate reduces the one-particle evolution to:
  \begin{equation}
      \frac{\partial f(\vec{p},t)}{\partial t} = n(t)\Gamma(\vec{p}) - \frac{f(\vec{p},t)}{\tau_R}.
  \end{equation}
The stationary solution to this equation requires $\Gamma(\vec{p}) = f_{\rm eq}(\vec{p}) / (n\tau_R)$. Substituting this back precisely yields the rest-frame BGK equation:
\begin{equation}
    \frac{\partial f(\vec{p},t)}{\partial t} = -\frac{1}{\tau_R}\bigl(f(\vec{p},t) - f_{\rm eq}(\vec{p})\bigr).
\end{equation}
  
To write this in a frame-independent, Lorentz-covariant form, the time derivative is promoted to the covariant derivative along particle worldlines. In a frame moving with fluid four-velocity $u^\mu$, the rest-frame particle energy $p^0$ corresponds to the local rest frame energy, $-p^\mu u_\mu$. The unique covariant generalization that preserves a momentum-independent relaxation time in the local rest frame is the Anderson-Witting equation:
\begin{equation}
   p^\mu \partial_\mu f = -\frac{p^\mu u_\mu}{\tau_R}(f - f_{\rm eq}).
\end{equation}

\subsection{Bjorken Flow and the Universal Attractor}

To validate the relaxation kernel's ability to capture the non-equilibrium attractor independently of spatial advection artifacts, we simulate a $0+1$D transversely homogeneous Bjorken expansion~\cite{Bjorken:1982qr} using a static computational box~\cite{El:2007vg, El:2008yy}. In this approach, the fluid remains macroscopically at rest, and the longitudinal expansion is enacted microscopically via a geometric redshift of the longitudinal momentum during the advection step:
\begin{equation}
    p_z(\tau + \Delta \tau) = p_z(\tau) \frac{\tau}{\tau + \Delta \tau}.
\end{equation}

The local energy density scales accordingly by $\tau_0/\tau$. This isolates the stochastic relaxation kernel, forcing a dynamical competition between the macroscopic geometric expansion (which drives the system toward a pressure anisotropy $P_L \ll P_T$) and the microscopic RAMBO redraws (which drive the system toward local isotropy, $P_L = P_T$).

For an ideal fluid ($\eta/s \to 0$), this expansion yields the exact cooling rate $T \propto \tau^{-1/3}$. At finite $\eta/s$, viscous heating slows this cooling. More fundamentally, the system must lose memory of its initial state and collapse onto a universal non-equilibrium attractor~\cite{Romatschke:2017vte,  Strickland:2018ayk, Strickland:2019hff, Jaiswal:2019cju}. To evaluate this, we track the scaled shear stress $\bar{\pi}(\tau) = (\epsilon/3 - T^{zz})/(4\epsilon/3)$ against the dimensionless conformal time $\tilde{w} \equiv \tau T / (5\eta/s)$. 

\begin{figure}[t]
\centering
\includegraphics[width=\linewidth]{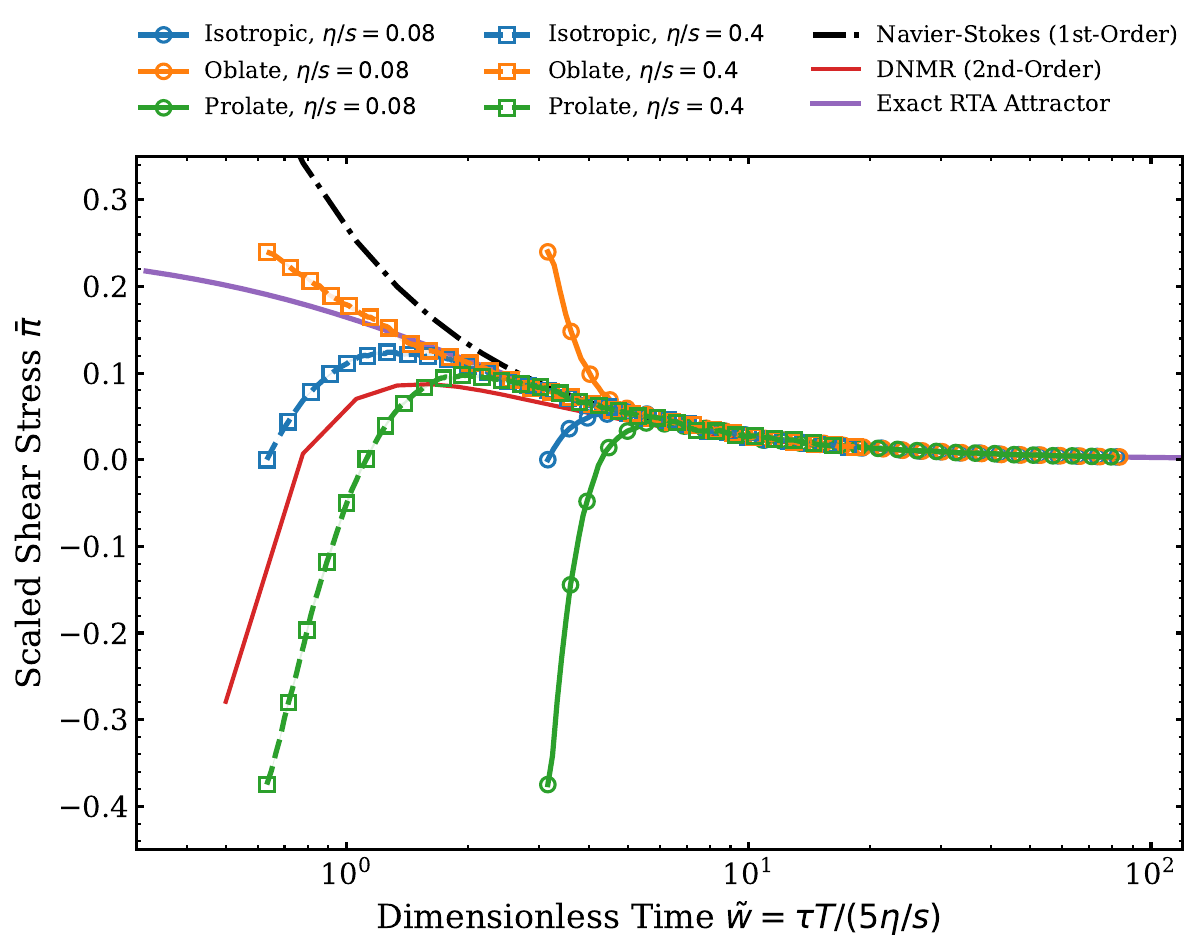}
\caption{Evolution of the inverse Reynolds number $\bar{\pi} = [(1/3) \epsilon - T_{zz}]/[(4/3) \epsilon]$ as a function of the dimensionless universal time $\tilde{w} = \tau T/(5\eta/s)$ for a Bjorken expansion. Numerical results from ShARK are shown for varying initial momentum anisotropies (isotropic, oblate, and prolate) and macroscopic transport coefficients ($\eta/s = 0.08$ and $0.4$). These are compared against the first-order Navier-Stokes limit, the second-order DNMR approximation, and the exact all-order RTA kinetic attractor of Ref.~\cite{Strickland:2018ayk}.}
\label{fig:bjorken_attractor}
\end{figure}

We initialized simulations in three disparate microscopic configurations: an isotropic state, an extreme oblate deformation, and an extreme prolate deformation. As shown in~Fig.~\ref{fig:bjorken_attractor}, despite these varied initial conditions, all trajectories rapidly converge onto a single universal curve at $\tilde{w} \approx 2$.

As an independent check that this collapsed trajectory reproduces the true all-order kinetic attractor, we benchmarked the ShARK output against the exact semi-analytic solution of the RTA Boltzmann equation for Bjorken flow~\cite{Baym:1984np, Florkowski:2013lya}, expressed as a Volterra integral equation for the energy density:
\begin{multline}
    \epsilon(\tau) = \Lambda_0^4 D(\tau, \tau_0) R(\xi(\tau, \tau_0)) + \\
    \int_{\tau_0}^\tau \frac{d\tau^\prime}{\tau_R} D(\tau, \tau^\prime) \epsilon(\tau^\prime) R(\xi(\tau, \tau^\prime))
\end{multline} 
where $D(\tau_2, \tau_1)$ is the exponential damping function and $R(z)$ is the free-streaming momentum anisotropy kernel. Evaluating this integral from an initial free-streaming limit ($\xi_0 \gg 1$ at $\tau_0 \ll 1$) yields the exact theoretical attractor of Refs.~\cite{Romatschke:2017vte, Strickland:2018ayk, Jaiswal:2019cju}.
 
\bibliography{ref}
 
\end{document}